\documentclass[article]{article}
\usepackage{graphics}
\usepackage{graphicx}
\usepackage{enumerate}

\begin{document}

\title{Analytical evaluation of the numerical coefficients in the Zassenhaus product formula and its applications to quantum and statistical mechanics}

\author{Mauro Bologna }

\date{ Instituto de Alta Investigaci\'{o}n,
Universidad de
  Tarapac\'{a}-Casilla 7-D Arica, Chile.}

 \maketitle



\abstract {This paper studies the exponential of the sum of two non-commuting operators as an infinite product of exponential operators involving repeated commutators of increasing order. It will be shown how to determine two coefficients in front of the nested commutators in the Zassenhaus formula. The knowledge of one coefficient is enough to generate a closed formula that has several applications in solving problems ranging from linear differential equations, quantum mechanics to non-linear differential equations.}

\section{Introduction}\label{secintro}
A crucial problem of mathematical physics is the development of the exponential
of the sum of two non-commuting operators. A very deep and important application of the Baker-Campbell-Hausdorff
formula~\cite{camp,baker,hausd} and its dual, the Zassenhaus formula~\cite{magnus,que,casas}, is related to quantum mechanics. Quantum mechanics is the kingdom of the operators and the Baker-Campbell-Hausdorff formula, jointly with the Zassenhaus formula, are an important mathematical tool. Even the simplest case of a time-independent one-dimensional Hamiltonian shows the difficulty of the problem. If considered by a general point of view, both the time evolution of a state~$|\psi(t)\rangle$ and the eigenvalues problem associated with it are formidable problems. Indeed the following well-known Schr\"odinger's equations

\begin{eqnarray}\label{eqham}
&i\hbar \frac{\partial}{\partial t}|\psi(t)\rangle=H| \psi(t)\rangle,\,\,H\psi(x)=E\psi(x),\\\label{eqham2}
&H\psi(x)=E\psi(x),\,\,H=-\hbar^2 \frac{\partial^2}{\partial x^2}+V(x),
\end{eqnarray}
where~$V(x)$ is a generic potential, are an unsolved problem. Exact solutions are known for a particular choice of the potential function only. Schr\"odinger's equations is a particular case of a linear differential equation of order~$n$ and with variable coefficients
\begin{equation}\label{eq1}
y^{(n)}(z)+a_{n-1}(z)y^{(n-1)}(z)+\cdots +a_0(z)y(z)=0,
\end{equation}
where~$y^{(j)}$ means the~$j$th derivative of~$y(z)$. It can be rewritten as
 
 \begin{equation}\label{eq1matrix}
 \frac{d}{dz}Y(z)=Y(z)M(z), \,\,Y(z)\equiv\left [y(z)\cdots,y^{(n-1)}(z)\right].
\end{equation}
Reducing Eq.~(\ref{eq1}) to  a linear differential equation for a linear operator, Eq.~(\ref{eq1matrix}), then we may write a formal solution via the Magnus expansion~\cite{magnus}. This last case is directly related to the solution of the problem of a time-dependent Hamiltonian that is an intensive research field~\cite{marini,kim,cerv,pedro,zurek,appl1,bader,zass3,bader2,zass4}. In particular in Refs.~\cite{bader,bader2} the investigation is performed using the Zassenhaus formula.  In the present paper, we will find an approximated formula obtained from the Zassenhaus expansion based on the knowledge of the coefficients in front of the nested commutators.
 
Another important active research field is Statistical Mechanics. We will focus on the liouvillian approach to statistical mechanics, i.e.

\begin{eqnarray}\label{liou_start}
\frac{\partial}{\partial t}\rho(q_1\cdots q_n,t)=-\sum_{i=0}^n \frac{\partial}{\partial q_i} [a_i(q_1\cdots q_n,t)\rho(x,t)].
\end{eqnarray}
The associated Langevin equations (Ref.~\cite{don} for an historical point of view and~\cite{peter} for a generalization) are

\begin{eqnarray}\label{nnlin_start}
\frac{dq_i}{dt}=a_i(q_1\cdots q_n,t),\,\,i=1,\cdots n,
\end{eqnarray}
where~$a_i(q_1\cdots q_n,t)$ are given functions that in general can be  stochastic. Its connection to statistical mechanics can be found via the Van Kampen's lemma~\cite{van} that relates the liouvillian density to the probability density through the relation~$P(\mathbf{q},t)=\langle\rho(\mathbf{q},t)\rangle$. It is virtually impossible to list a complete bibliography on the
topic. For this reason, the reader is referred to few exemplary textbooks or paper collections~\cite{west_app,west_coll2,west_coll3,west_coll4,pet3}. 

As we may infer by direct inspection, Eqs.~(\ref{nnlin_start}) are a set of non-linear equations whose solution is generally unknown. Using the results of the paper we will find an approximate analytical solution of Eqs.~(\ref{nnlin_start}) and consequently of Eq.~(\ref{liou_start}). To realize that we will use the Liouville equation or more in general, a linear partial differential equation. Solving non-linear problems through the solution of linear problems is a technique used in literature (see Ref.~\cite{bruce} on this topic) and a recent method based on spectral decomposition is presented in Ref.~\cite{gos}. Applying the results of the paper we will be able to evaluate the argument of the liouvillian density and through the method of characteristics, we will give a formula for the solution of Eqs.~(\ref{nnlin_start}).

The paper is organized as follows: In Sec.~\ref{secondsec} we will obtain an analytical expression for the coefficients of the nested commutators~$[A,[A,\cdots,[A,B]]]$ and~$[[[A,B],\cdots,B,B]$. In Sec.~\ref{secorddiff}, using the result of the previous section, we will find an approximated closed formula for linear differential equations. In Sec.~\ref{ivsecmechquantum} we will apply the previous results to recover the fundamental energy level of a quantum system in a generic shallow potential well and we will shortly sketch a solution for the time evolution of a quantum state. In Sec.~\ref{vstatisticalmech}, using the result of Sec.~\ref{secondsec}, we will find an approximated closed formula for the first order non-linear differential equations applied to statistical mechanics.

\section{The Zassenhaus product formula}\label{secondsec}
In this section, we will study an analytical approach to determine the coefficients in the Zassenhaus exponents in front of the nested commutators. The knowledge of these coefficients inspired several recent works (see for example ~\cite{zass1m,zass2m,zass1}). As we will see, the result will have several important applications. Our starting point is the analytical solution of the operatorial differential equation obtained in Ref.~\cite{yo}. Given the equation  

\begin{equation}\label{eq2}
\frac{d}{dt}Y(t)=M(t)Y(t),
\end{equation}
its solution is  

\begin{equation}\label{fmatr7}
Y(t)=\left[I+\int_0^tM(t_0)\exp\left[u
O\right]\mid_{t_0=0}du\right]Y(0)
\end{equation}
where~$M(t)$ is a time dependent operator, such as for example a~$n\times n$ matrix, and~$O$ is defined as

\begin{equation}\label{o}
O\equiv \frac{d}{dt}+M(t).
\end{equation}
The symbol~$I$ is the identity matrix and it is understood that~$O$ acts on the left, i.e. on~$M(t_0)$. The symbol~$\exp\left[u O\right]M(t_0)\mid_{t_0=0}$ means that it must be evaluated at~$t_0=0$. Equivalently for the equation

\begin{equation}\label{eq3}
\frac{d}{dt}Y(t)=Y(t)M(t)
\end{equation}
we have

\begin{equation}\label{fmatr6}
Y(t)= Y(0)\left[I+\int_0^t\exp\left[u
O\right]M(t_0)\mid_{t_0=0}du\right]
\end{equation}
where it is understood that~$O$ acts on the right. The aim of this section is to rigorously show that the coefficient in the Zassenhaus product formula associated to the term $[A,[A,\cdots,[A,B]]]$ is~$(-1)^{n}/n!$ while $[[[A,B],\cdots,B,B]$ is~$-(n-1)/n!$. 

As starting point, we will prove that when the matrix~$M(t)$ and its derivatives commute at any value of the parameter~$t$ than we will recover the solution of the differential equation (\ref{eq2}). To fix the ideas let us assume that~$M(t)$ is a diagonal matrix. Using the Trotter product formula~\cite{nelson} we may rewrite solution (\ref{fmatr6}) as

\begin{eqnarray}\nonumber
&&Y(t)=Y_0\left(I+
\lim_{N\to\infty}\prod_{j=1}^{N}\int\limits_0^t
\exp\left[\frac{u}{N}
\partial_{t_0}\right]\!\cdots\right.\\
&&\left.\exp\left[\frac{u}{N}
 M(t_0)\right]\!M(t_0)\mid_{t_0=0}\!du\!\right).
\end{eqnarray}
Note that the operator~$\exp\left[\frac{u}{N}
\partial_{t_0}\right]$ translates of a quantity~$\frac{u}{N}$ the
argument of ~$M(t_0)$. Without loss of generality we set~$t_0=0$.  Performing the translation and exploiting the commutating properties of the matrix~$M$  we may write

\begin{eqnarray}\label{fmatr10}
Y(t)=Y_0\left(I+
\lim_{N\to\infty}\int\limits_0^t\prod_{j=1}^{N}
\exp\left[\frac{u}{N}
M\left(j\frac{u}{N}\right)\right] M\left(u\right)du\right).
\end{eqnarray}
Since~$M\left(j\frac{u}{N}\right)$ commutes for any value of~$j\frac{u}{N}$, we rewrite Eq.~(\ref{fmatr10}) as

\begin{eqnarray}\label{fmatr10b}
Y(t)=Y_0\left(I+
\lim_{N\to\infty}\int\limits_0^t
\exp\left[\sum_{j=1}^{N}\frac{u}{N}
M\left(j\frac{u}{N}\right)\right]M\left(u\right) du\right).
\end{eqnarray}
We now use the fact that for~$N\to\infty$ we can write the sum in the continuous limit, i.e.

\begin{eqnarray}\label{limite}
\frac{u}{N}\sum_{j=1}^{N}
 M\left(j\frac{u}{N}\right)\to\int\limits_0^u
 M\left(z\right)dz.
\end{eqnarray}
Plugging the result into Eq.~(\ref{fmatr10b}) we finally obtain

\begin{eqnarray}\label{soluzione}
Y(t)\!=\!Y_0\!\left(\!I\!+\!
\int\limits_0^t\!
\exp\left[\!\int\limits_0^u
 M\left(z\right)dz\!\right]\! M\left(u\right)du\right)\!=
\!Y_0\exp\left[\!\int\limits_0^t
 M\left(z\right)dz\!\right]\!.
\end{eqnarray}
We will use this result to determine the coefficient in front of the term $[A,[A,\cdots,[A,B]]]$ in the Zassenhaus expansion formula. Indeed if we consider the Zassenhaus expansion formula~\cite{magnus}

\begin{eqnarray}\nonumber
&& \exp[u(A+B)]=\exp\left[uA\right]\exp\left[u B\right]
\exp\left[-\frac{1}{2}u^{2}\left[A,B\right]\right] \\\label{exp}
&&\exp\left[\frac{1}{6}u^{3}[A,[A,B]]-
\frac{1}{3}u^{3}[[A,B],B]\right]\cdots
\end{eqnarray}
and setting~$A=\partial_t$ and~$B=M(t)$ we have that

\begin{eqnarray}\label{comm}
 [\partial_t,M(t)]=\partial_t M(t)\equiv M^{(1)},\cdots,[\partial_t,[\partial_t,\cdots,[\partial_t,M(t)]]]=M^{(n)}.
\end{eqnarray}
Formally solution of Eq.~(\ref{fmatr6}) can be rewritten as

\begin{eqnarray}\nonumber
&&Y(t)=Y_0\left(I+
 \int\limits_0^t\exp\left[u\partial_t\right]\exp\left[u M(t)\right]
 \exp\left[-\frac{1}{2}u^{2}\left[\partial_t,M(t)\right]\right]
\right.\\\nonumber
&&\left.\exp\left[\frac{1}{6}u^{3}[\partial_t,[\partial_t,M(t)]]-
\frac{1}{3}u^{3}[[\partial_t,M(t)],M(t)]\right]\right.
\\\label{zass}
&&\left.\cdots M\left(t\right)\mid_{t=0}du\right).
\end{eqnarray}
Since~$M(t)$ and its derivatives are commutating matrices~$\forall t$, all the commutators present in the formula vanish except for those that contain only once the matrix~$M$. Using the properties of the translational operator~$\exp\left[u\partial_t\right]$ we may write the following

\begin{eqnarray}\nonumber
&&Y(t)=Y_0\!\left(\!I\!+\!
\! \int\limits_0^t\!\exp\left[u M(u)\right]
\exp\left[-\frac{1}{2}u^{2}M^{(1)}(u)\right]\exp\left[\frac{1}{6}u^{3}M^{(2)}(u)\right]\right.\\\label{zass2}
&&\left.\cdots\exp\left[c_nu^{n}M^{(n-1)}(u)\right]\cdots M\left(u\right) du\right).
\end{eqnarray}
Exploiting again the commutating properties of~$M(t)$, the exponential matrix product is commutative and we may write Eq.~(\ref{zass2}) as

\begin{eqnarray}\label{zass3}
Y(t)=Y_0\left(I+
 \int\limits_0^t\exp\left[\sum_{n=1}^{\infty}c_nu^{n} M^{(n-1)}(u)\right]
M\left(u\right) du\right).
\end{eqnarray}
This result has to be the same result obtained in Eq.~(\ref{soluzione}). This is so if and only if

\begin{eqnarray}\label{zass4}
\int\limits_0^u
 M\left(z\right)dz=\sum_{n=1}^{\infty}c_nu^{n} M^{(n-1)}(u)
\end{eqnarray}
Integrating by part repeatedly the left side of Eq.~(\ref{zass4}) we end up into

\begin{eqnarray}\label{zass5}
\sum_{n=1}^{\infty}(-1)^{n-1}\frac{u^{n}}{n!} M^{(n-1)}(u)=\sum_{n=1}^{\infty}c_nu^{n} M^{(n-1)}(u).
\end{eqnarray}
Equating the coefficients of the power laws of the two series we infer that

\begin{eqnarray}\label{zassc}
 c_n=\frac{(-1)^{n-1}}{n!}.
\end{eqnarray}
We may use the same development in terms of nested commutators identifying~$A$ with~$M$ and~$B$ with~$\partial_t$. In this case Eq.~(\ref{zass}) may be written as

\begin{eqnarray}\nonumber
&&Y(t)=Y_0\left(I+
 \int\limits_0^t\exp\left[u M(t_0)\right]\exp\left[u\partial_t\right]
 \exp\left[-\frac{1}{2}u^{2}\left[M(t),\partial_t\right]\right]
\right.\\\nonumber
&&\left.\exp\left[\frac{1}{6}u^{3}[M(t),[M(t),\partial_t]]-
\frac{1}{3}u^{3}[[M(t),\partial_t],\partial_t]\right]\right.
\\\label{zass6}
&&\left.\cdots M\left(t\right)\mid_{t=0}du\right).
\end{eqnarray}
As before the solution has to be the solution given in Eq.~(\ref{soluzione}). But in this case the identification of the correspondent coefficient is not so straightforward. We assume the same conditions on the matrix~$M$ as before but we consider that its elements are polynomial. Without loss of generality, we set its matrix element as~$f(t)=t^{n}$. It is no hard to see that the~$m$-th nested commutator is~$\left[\left[M(t),\partial_t\right]\cdots\right]=(-1)^{m}M^{(m)}(u)$. Since the matrices commute, we may perform the sum of the exponentials~$\exp\left[c_m \left[\left[M(t),\partial_t\right]\cdots\partial_t\right]\right]$ and, according to Eq.~(\ref{zass4}), we must have

\begin{eqnarray}\label{zass7}
 \sum _{m=0}^{n}\bar{c}_m\frac{(-1)^{m}n!}{(n-m)!}u^{n+1}= \int_0^uM_{ij}(z)dz =\frac{1}{n+1}u^{n+1}.
\end{eqnarray}
For each power of the polynomial we determine the coefficient~$\bar{c}_m$. We may write the following recursive relationship

\begin{eqnarray}\label{zass8}
 \bar{c}_n=\frac{(-1)^n}{(n+1)!}-\sum _{m=0}^{n-1}\bar{c}_m\frac{(-1)^{n-m}}{(n-m)!}.
\end{eqnarray}
It is straightforward to show that Eq.~(\ref{zass8}) correspond to

\begin{eqnarray}\label{zass9}
 \bar{c}_n=-\frac{(n-1)}{n!}.
\end{eqnarray}
Formulas (\ref{zassc}) and (\ref{zass9}) give an analytical expression for the coefficients in front of the nested commutator~$[A,[A,\cdots,[A,B]]]$ and~$[[[A,B],\cdots,B,B]$ respectively. The exactness of those formulas can be checked in Ref.~\cite{zass1m,zass2m,zass1}.

\section{Ordinary differential equations with small variable coefficients}\label{secorddiff}
In this section, we will use the result of Sec. \ref{secondsec} to provide an approximate solution of a differential equation in the form

\begin{eqnarray}\label{eigm1}
y^{(n)}(t)+a_{n-1}(t,\varepsilon)y^{(n-1)}(t)+\cdots +a_0(t,\varepsilon)y(t)=0
\end{eqnarray}
where~$\varepsilon$ is a small parameter and~$a_{n-1}(t,\varepsilon)\to 0$ for~$\varepsilon\to 0$. The above equation can be written as a first order differential equation in the form of Eq.~(\ref{eq3}). We will consider the problem set in Eq.~(\ref{eigm1}) when the matrix~$M$ may be written as~$M(t)=\varepsilon N(t)$, where~$N(t)$ is a matrix with finite values of its elements. We assume that it holds the following equality

\begin{eqnarray}\label{eigm4}
\left[M^{(s)}(t),M^{(p)}(t)\right]=0,\,\, \textrm{for}\,\,s,p\geq k
\end{eqnarray}
where~$k$ is a given integer. Using the results of the previous section we obtain at~$\varepsilon$ order

\begin{eqnarray}\nonumber
&&Y(t)\approx Y_0\left(I+\int_0^t\exp[u M(u)]\cdots \exp\left[c_{k}u^{k} M^{(k-1)}(u)\right]\right.\\\nonumber
&&\left.\exp\left[\sum_{n=k+1}^{\infty}c_{n}u^{n} M^{(n-1)}(u)\right]M(u)du\right)\!=\!
\\\nonumber
&&=Y_0\!\left(\!I+\int_0^t\!\!\exp[u M(u)]\cdots \exp\left[c_{k}u^{k} M^{(k-1)}(u)\right]\right.\\\label{eigm5}
&&\left.\exp\!\!\left[\int_{0}^{u}\!\!M(z)dz-\sum_{n=1}^{k}c_{n}u^{n} M^{(n-1)}(u)\!\right]\!M(u)du\!\right)
\end{eqnarray}
where the coefficients~$c_n$ are given by Eq.~(\ref{zassc}). In particular if~$M(t)$ commutates starting from the second derivative on, we have the following approximate formula

\begin{eqnarray}\label{eigmx}
Y(t)\!\approx \!Y_0\left(\!I+\int_0^t\!\exp[u M(u)]\exp\left[\int _0^u\!M(z)dz-u M(u)\right]M(u) du \!\right).
\end{eqnarray}
As example we consider the following second order differential equation

\begin{eqnarray}\label{eigmx3}
\frac{d^{2}}{dt^{2}}x(t)=\varepsilon^{2}f(t)x(t).
\end{eqnarray}
Note that the small coefficient~$\varepsilon$ can be eliminated by rescaling the variable~$t$ and leading to an alternative formulation of the problem

\begin{eqnarray}\label{eigmx3_al}
\frac{d^{2}}{d\tau^{2}}x(\tau)=f(\tau/\varepsilon)x(\tau).
\end{eqnarray}
The above equations can be generated by a two dimensional system

\begin{eqnarray}\label{eigmx3b}
\frac{d}{dt}[x(t),y(t)]=[x(t),y(t)] M(t),\,\,\,\,
M(t)=\varepsilon\left[
\begin{array}{cc}
 0 & f(t) \\
1& 0\\
 \end{array}
 \right],
\end{eqnarray}
and, applying Eq.~(\ref{eigmx}), we obtain

\begin{eqnarray}\nonumber
x(t)\!&=&\!a\! \left(\!1+\!\int_0^t\! \varepsilon^{2}\Delta(u)\cosh\! \left[u \varepsilon  \sqrt{\!f(u)}\right]+\varepsilon\sqrt{\!f(u)} \sinh\! \left[u \varepsilon  \sqrt{f(u)}\right]\!du\right)\!+\\\label{eigmx4}&+&b\varepsilon \int_0^t \left(\frac{\varepsilon \Delta(u)\sinh \left[\varepsilon  u \sqrt{f(u)}\right]}{\sqrt{f(u)}}+\cosh \left[\varepsilon  u \sqrt{f(u)}\right]\right)du,\\\nonumber y(t)&=&a\varepsilon \int_0^t f(u) \cosh\left[\varepsilon  u \sqrt{f(u)}\right]du+\\\label{eigmx5}
&&b \left(1+\varepsilon\int_0^t
\sqrt{f(u)} \sinh\left[\varepsilon u \sqrt{f(u)}\right]du\right),
\end{eqnarray}
where

\begin{eqnarray}\label{eigmx6}
\Delta(u)= \int_0^u f(z) \, dz-u f(u).
\end{eqnarray}
It is no hard to see that a sufficient condition so that Eqs. (\ref{eigmx4}) and (\ref{eigmx5}) are a valid approximation in~$t\in [a,b]$ is 

\begin{eqnarray}\label{condi}
\left|  \int_0^t f(z) \, dz-t f(t)\right|=\mid \Delta(t)\mid\le K
\end{eqnarray}
where~$K$ is a finite positive number. For an infinite interval,~$t\in [0,\infty)$, a sufficient condition is given by~$f(t)=k +g(t)$ where~$k$ is a finite value parameter and~$g(t)$ is a finite value function such that~$g(t)\to 0$ faster than~$1/t$ for~$t\to\infty$. Less strict conditions can be found according to the case under study. As example, we consider the differential equation

\begin{eqnarray}\label{exsimpl1}
 \frac{d^{2}}{dt^{2}}x(t)=-\varepsilon^{2}\left(1-t\exp\left[-t^2\right] \right)x(t),\,\,x(0)=1,\,\,
\left.\frac{dx}{dt}\right\vert_{t=0}=0,\,\,\varepsilon=0.2.
 \end{eqnarray}
The coefficient~$f(t)$ is on purpose taken with an exponentially decaying term. This will show that the contribution of such a term to the correction to the unperturbed solution~$(x_0(t),x_0'(t))=(\cos\varepsilon t, -\varepsilon\sin\varepsilon t)$ is not negligible. The approximate solution given by Eq.~(\ref{eigmx4}) is visually indistinguishable and it is more convenient to plot the percent error. Since the solution is oscillating, to plot the percent error we can not use the traditional definition~$\mid x_{ex}(t)-x_{app}(t)\mid/\mid x_{ex}(t)\mid$. This because when the exact solution~$ x_{ex}(t)$ vanishes, the error would be divergent, regardless of how the approximate value is near to the exact one. To overcome this difficulty we first define the percent error as the "distance" between two functions with respect one of the two. In mathematical symbols

\begin{eqnarray}\label{norm}
 \Delta E=100\frac{\| y_{2}(t)-y_{1}(t)\|_{L_1}}{\| y_{2}(t)\|_{L_1}} 
 \end{eqnarray}
where~$y_{2}(t)$ is the reference function,~$ y_{1}(t)$ is the approximated function and~$\| \cdot\|_{L_1}$ is defined as

\begin{eqnarray}\label{norm2}
 \| y(t)\|_{L_1}=\int_I |y(t)|dt
 \end{eqnarray}
 
\begin{figure}[ht]
\begin{minipage}[b]{0.45\linewidth}
\centering
\includegraphics[width=.8\textwidth]{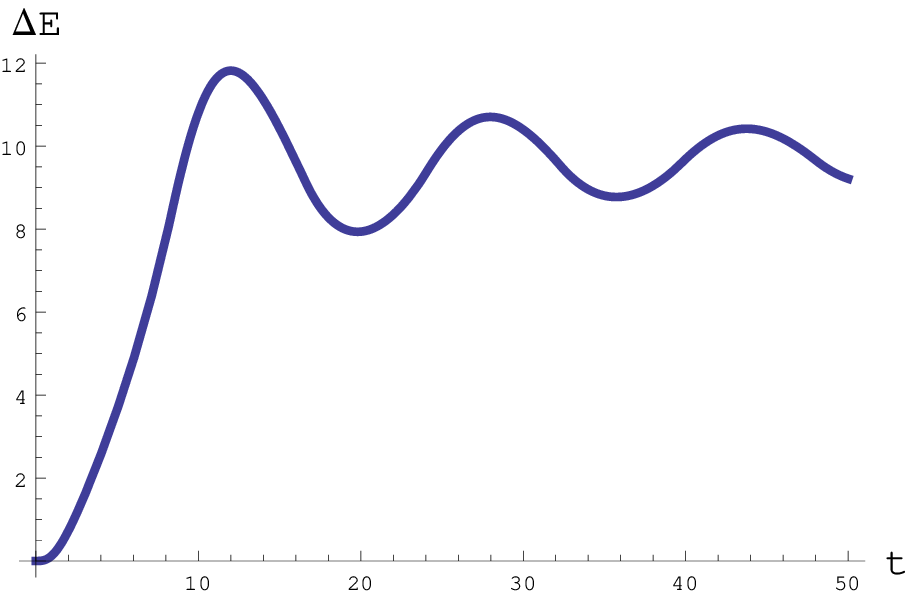}
\caption{Plot of the the percent error~$\Delta E$. On the horizontal axis the~$t$ variable while the vertical axis represents~$\Delta E$. Here~$x_{app}(t)=\cos\varepsilon t$.
\label{contr}}
\end{minipage}
\hspace{0.1cm}
\begin{minipage}[b]{0.45\linewidth}
\centering
\includegraphics[width=.8\textwidth]{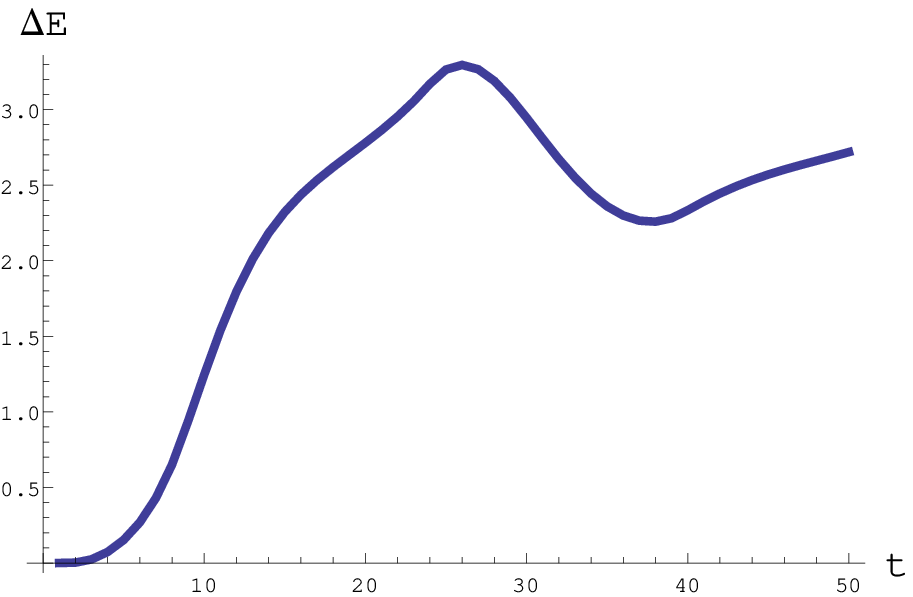}
\caption{Plot of the the percent error~$\Delta E$. On the horizontal axis the~$t$ variable while the vertical axis represents~$\Delta E$. Here~$x_{app}(t)$ is given by Eq.~(\ref{eigmx4}) with~$a=1$ and~$b=0$.
\label{del}}
\end{minipage}
\end{figure}
In Fig.~\ref{contr} it has been plotted the percent error between the exact solution and the "unperturbed" solution i.e.~$x_{0}(t)=\cos\varepsilon t$. The percent error, given by~$ \| x_{ex}(t)-x_{0}(t)\|_{L_1}/\| x_{ex}(t)\|_{L_1} 100$, is of the order of~$10\%$. In Fig.~\ref{del} it has been plotted the percent error between the exact solution and~$x(t)$ given by Eq.~(\ref{eigmx4}). The percent error is of the order of~$2\%$. We end the section noting that if we consider the derivative of the solution then the percent error between the exact solution and the unperturbed solution~$x_0'(t)=-\varepsilon\sin\varepsilon t$ ranges from~$10\%$ up to~$50\%$. Taking the derivative of Eq.~(\ref{eigmx4}), the percent error between the exact solution and the approximate solution is of the order of~$1\%$. Finally, we note that one can build more accurate solutions rearranging~$x(t)$ the derivative of the function~$x(t)$ and~$y(t)$ according to the initial values~$(x_0,y_0)$. A detailed study of this topic is out of the scope of this paper.
 
\section{Applications to quantum mechanics: shallow potential well and time evolution of a quantum state}\label{ivsecmechquantum}
\begin{figure}[ht]
\centering
\includegraphics[width=.5\textwidth]{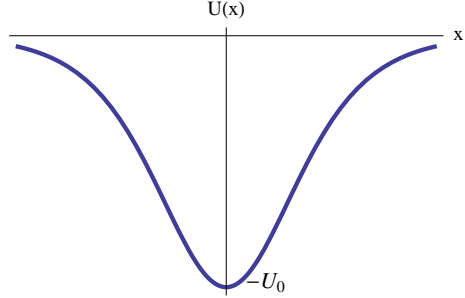}
\caption{The plot represents  a potential well~$V(x)=-V_0g(x/a)$ of depth~$V_0$ .
\label{well}}
\end{figure}

Many works in quantum mechanics are devoted to finding approximate solutions of Schr\"odinger's equation. Among them, several methods use Zassenhaus expansion formula~\cite{bader,bader2}. As already noticed in the previous sections, this paper is more focused on the wide range of possible applications of the proposed method rather than its accuracy. In spite of the fact that we are working with a first order approximation, the accuracy is enough to find both the energy level of a quantum particle in a shallow potential well and the time evolution of a quantum state. First we will study the stationary Schr\"odinger's equation to evaluate the energy of a particle in a shallow potential well. The one dimension Schr\"odinger's equation reads as

\begin{eqnarray}\label{qm1}
\frac{d^{2} \psi}{dx^{2}}+\frac{2m}{\hbar^{2}}\left[E-V(x)\right]\psi=0
\end{eqnarray}
where for shortness we dropped off the argument of the wave function. Without loss of generality, we assume that the potential is of the form~$V(x)= -V_0 g(x/a)$ where~$a$ is a length scale parameter and~$g(z)$ a positive function. Performing the variable change~$z= x/a$ and defining~$e=E/V_0$ we may rewrite Eq.~(\ref{qm1}) as

\begin{eqnarray}\label{qm2}
\frac{d^{2} \psi}{dz^{2}}+\varepsilon^{2}\left[e+g(z)\right]\psi=0,\,\,\,\, 
\varepsilon^{2}\equiv \frac{2mV_0 a^2}{\hbar^{2}}.
\end{eqnarray}
We will focus on a shallow potential well. The literature on shallow potential well is vast and we limit ourselves to Ref.~\cite{landau3} as textbook.  In particular, we consider a symmetric potential with a finite value for the minimum as shown in Fig.~\ref{well}. 
The condition for shallow potential well is 
$\varepsilon^{2}\ll 1$. Since we are interested to the bound state, i.e.~$e=-|e|<0$ with~$g(x)>0$, we rewrite Eq.~(\ref{qm2}) as

\begin{eqnarray}\label{qm3}
\frac{d^{2} \psi}{dz^{2}}-\varepsilon^{2}\left[|e|-g(z)\right]\psi=0.
\end{eqnarray}
Setting~$f(z)=|e|-g(z)$, if~$g(z)\to 0$ faster than~$1/|z|$ for~$|z|\to\infty$, then condition (\ref{condi}) applies. Vanishing the coefficient of the positive exponentials in solutions (\ref{eigmx4}) and (\ref{eigmx5}),  we obtain the necessary conditions to have a decaying solution at infinity. At the zero and first order on~$g(z)$, we have

\begin{eqnarray}\label{qm4}
\left(-\varepsilon \Delta_{\infty} +\sqrt{|e|}\right) a+\left(1-\varepsilon \frac{\Delta_{\infty}}{\sqrt{|e|}}\right)b=0,\,\,\,\,\Delta_{\infty} \equiv \int_0^{\infty}g(z)dz
\\\label{qm5}
\sqrt{|e|} a+ \frac{\varepsilon \Delta_{\infty} }{\sqrt{|e|}}b=0.
\end{eqnarray}
Vanishing the determinant of the above system we have the condition

\begin{eqnarray}\label{qm6}
 \left(-\varepsilon \Delta_{\infty} +\sqrt{|e|}\right)^2=0, \,\,\Rightarrow |e|=\varepsilon^2\Delta_{\infty}^2,
\end{eqnarray}
or, in terms of the energy~$E$,

\begin{eqnarray}\label{qm7}
E=-\frac{2mV_0^2 a^2}{\hbar^{2}}\Delta_{\infty}^2=
-\frac{m}{2\hbar^{2}}\left(\int\limits_{-\infty}^{\infty}V(x)dx\right)^2.
\end{eqnarray}
We recovered the result given at pag.~163 in Ref.~\cite{landau3}. 

We end this section shortly studying the time evolution of a quantum state. To apply directly the formalism of Sec.~\ref{secondsec} we consider the time evolution of~$\langle\psi(t)|$. Setting for brevity~$\hbar=1$, we have

\begin{equation}\label{eqham_time}
-i\frac{\partial}{\partial t} \langle\psi(t)|=\langle\psi(t)|(H_0+\varepsilon H_1(t))| 
\end{equation}
where~$H_0$ is a time-independent Hamiltonian and~$\varepsilon H_1(t)$ is a small time-dependent correction to~$H_0$. Passing to the interaction representation~\cite{landau4} we may always transform Eq.~(\ref{eqham_time}) into

\begin{equation}\label{eqham_time_bis}
-i \frac{\partial}{\partial t}\langle\psi_I(t)| =\langle\psi_I(t)| \varepsilon H_I(t)
\end{equation}
with~$ H_I(t)=\exp[i H_0 t] H_1(t)\exp[-iH_0t]$. Using the results of Sec.~\ref{secondsec} at~$\varepsilon$ order we may write the approximate solution

\begin{eqnarray} \label{eqham_time2}
\langle\psi_I(t)|\approx \langle\psi_I(0)|\left(I+\prod\limits_{n=1}^{\infty}\int_0^t
\exp\left[i\varepsilon c_nu^n H^{(n-1)}_I(t)\right] i \varepsilon H_I(u)du\right)
\end{eqnarray}
where the coefficient~$c_n$ are given by (\ref{zassc}). Without to specify the commutator~$[H^{(i)}_I(t),H^{(k)}_I(t)]$ no further calculations can be performed. Finally, we note that if the perturbation~$H_1(t)$ is time-independent, then the approximate solution for the time evolution of a state~$|\psi(t)\rangle$ is given by

\begin{eqnarray}\nonumber
&&|\psi(t)\rangle \approx   \exp\left[-i H_0t\right]\exp\left[-i\varepsilon H_1t\right]
\\\label{eqham_notime}  
&&\prod\limits_{n=2}^{\infty} 
\exp\left[-i\varepsilon c_nt^n [H_0,[H_0,\cdots,[H_0,H_1]]]\right]|\psi(0)\rangle.
\end{eqnarray}

\section{Statistical mechanics and non-linear equations}\label{vstatisticalmech}
\begin{figure}[ht]
\centering
 \includegraphics[width=.5\textwidth]{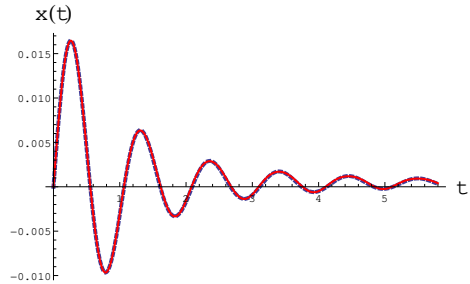}
\caption{The dotted line represents the numerical solution of Eq.~(\ref{liou3_cos}) while the continuous line is the plot of solution~(\ref{sol_cos}). The value of the parameters are:~$a=-1$,~$\beta=2$,~$f_0=1$,~$b=1$,~$ \Omega = 6$,~$ \tau = 1$,~$ \varepsilon = 0.1$ .
}
\label{nonlin}
\end{figure}

Generally speaking, a non-linear first order differential equation is an unsolved problem. There is not a general method to solve such a differential equation. One can use a recursive method such as the method of successive approximations (see for example Ref.~\cite{kelley}). These kind of methods are quite hard to handle from an analytical point of view. In this section, we will provide an approximate closed formula for first order non-linear differential equations. Let us consider the non-linear first order equation

\begin{eqnarray}\label{nnlin}
\frac{d\mathbf{x}}{dt}=\mathbf{a}(\mathbf{x},t).
\end{eqnarray}
where~$\mathbf{x}$ is a vector and~$\mathbf{a}(\mathbf{x},t)$ is a given vectorial function. Its connection to statistical mechanics can be found simply interpreting Eq.~(\ref{nnlin}) as Langevin equation~\cite{don,peter}.
The corresponding Liouville equation is

\begin{eqnarray}\label{liou}
\frac{\partial}{\partial t}\rho(\mathbf{x},t)=-\nabla\cdot [\mathbf{a}(\mathbf{x},t)\rho(x,t)].
\end{eqnarray}
If~$\mathbf{a}(\mathbf{x},t)$ is a stochastic function then, via the Van Kampen's lemma~\cite{van}, the liouvillian density is related to the probability density by the relation~$P(\mathbf{x},t)=\langle\rho(\mathbf{x},t)\rangle$. The difficulty to solve  Eq.~(\ref{nnlin}) is inherited by Eq.~(\ref{liou}) since also for Eq.~(\ref{liou}) there is not a general method for solving it. But, using physical arguments, if we interpret Eq.~(\ref{nnlin}) as the motion equation of a particle, then the solution of Eq.~(\ref{liou}) is

\begin{eqnarray}\label{liou2}
\rho(\mathbf{x},t)= \delta(\mathbf{x}-\mathbf{x}_P(t))
\end{eqnarray}
where~$\delta(z)$ is the Dirac delta function and~$\mathbf{x}_P(t)$ is the solution of Eq.~(\ref{nnlin}). An important advantage of considering Liouville equation, or, more in general, a linear partial differential equation, is that for the partial differential equation we may use a linear theory. For our purposes, we can exploit the theory developed in Sec.~\ref{secorddiff}. For sake of simplicity, we shall focus on a one-dimensional case with a(x, t) a real function. The one-dimensional case can be related to anomalous diffusion. In the past thirty years, anomalous diffusion has been intensively investigated~\cite{4,5,6,7,pet}. We will study the following non-linear equation~\cite{sancho2,sancho} 
\begin{eqnarray}\label{liou3_dim}
\frac{dx}{dt}=-\frac{\partial}{\partial x}V(x)+F(t)  
\end{eqnarray}
where, in a over-damped regime,~$V(x)$ represent a potential and ~$F(t)$ can be interpreted as a force and more in general as a stochastic force. The above equation can be always rewritten as

\begin{eqnarray}\label{liou3}
\frac{dx}{dt}=\varepsilon\left[-\frac{\partial}{\partial x}v(x)+f(t)\right]
\end{eqnarray}
where now~$t$,~$x$,~$v(x)$ and~$f(t)$ are dimensionless quantities and~$\varepsilon$ is a parameter that we will take small. Alternatively, making the variable change~$t\to\tau/\varepsilon$, we may consider also equations like

\begin{eqnarray}\label{liou3_alt}
\frac{dx}{d\tau}= -\frac{\partial}{\partial x}v(x)+f\left(\frac{\tau}{\varepsilon}\right).
\end{eqnarray}
The corresponding Liouville equation is

\begin{eqnarray}\label{liou4}
\frac{\partial}{\partial t}\rho(x,t)=\varepsilon\frac{\partial}{\partial x}[v'(x)\rho(x,t)]-\varepsilon f(t)\frac{\partial}{\partial x}[ \rho(x,t)].
\end{eqnarray}
Due to the formalism developed in Sec.~\ref{secorddiff} it is more convenient to study the equivalent equation

\begin{eqnarray}\label{liou5}
\frac{\partial}{\partial t}\Phi(x,t)=\varepsilon [v'(x)-f(t)]\frac{\partial}{\partial x}\Phi(x,t)\equiv \Phi(x,t) M .
\end{eqnarray}
Both equations have as solution an arbitrary function~$G(x,t)=\mathrm{const}$ where the arguments~$x,t$ are the solution of the characteristic equation, i.e. Eq.~(\ref{liou3}). The knowledge of the characteristic curve implies the knowledge of the function~$x_P(t)$ of Eq.~(\ref{liou2}) and ultimately the knowledge of the statistical function~$\rho(x,t)$. Following the prescription of Sec.~\ref{secondsec}, a formal solution of Eq.~(\ref{liou5}) via Eq.~(\ref{fmatr7}) is given by

\begin{eqnarray}\nonumber
\Phi(x,t)&=&\Phi(x,0)\left[1+\int_0^t du \exp\left[u\left(
\partial_{t_0}+\varepsilon \overleftarrow{\partial}\!\!_{x}[v'(x) -f(t_0)]\right)\right] \times\right.
\\\label{liou6}
&\times&\left.\varepsilon\frac{\overleftarrow{\partial}}{\partial x}\varepsilon [v'(x)-f(t_0)]|_{t_0=0}du\right]
\end{eqnarray}
with the understanding that the partial derivative with respect to~$x$ acts on the left (stressed by the arrow on top of the partial derivative symbol), while the partial derivative with respect to~$t$ acts on the right. Applying Eq.~(\ref{eigmx}) we obtain the approximate solution

\begin{eqnarray}\nonumber
&&\Phi(x,t)\approx \Phi(x,0) +\Phi(x,0)\int_0^t\exp\left[\varepsilon u  \overleftarrow{\partial}\!\!_{x} [v'(x)-f(u)]\right]\times
\\\label{eigmx_b}
&&\exp\left[ -\overleftarrow{\partial}\!\!_{x}\varepsilon\left(\int_0^u f(z)dz-u f(u)\right)\right]\frac{\overleftarrow{\partial}}{\partial x}\varepsilon [v'(x)-f(u)] du .
\end{eqnarray}
To perform the action of the exponential operators we make a change of variable as following 

\begin{eqnarray} 
w=\frac{1}{\varepsilon}\int_0^{x} \frac{1}{u'(y)-f(u)} dy,\,\,\,\,\Phi(x,0) = \bar{\Phi}(w(x))
\end{eqnarray}
and we have

\begin{eqnarray} \nonumber
&\Phi&\!\!\!\!(x,0)\exp\left[\varepsilon u  \overleftarrow{\partial}\!\!_{x} [v'(x)-f(u)]\right] \!\!\exp\!\!\left[ -\overleftarrow{\partial}\!\!_{x}\varepsilon\left(\int_0^u f(z) dz-u f(u)\right)\right]=
\\\label{action}
&=&\bar{\Phi}\left[w(x-\varepsilon \Delta(u) )+  u \right]
\end{eqnarray}
where we used the definition of~$\Delta(u)$ given in Eq.~(\ref{eigmx6}). We may rewrite Eq.~(\ref{eigmx_b}) as

\begin{eqnarray}\label{eigmx_c}
\Phi(x,t)&\approx& \Phi(x,0) +\int_0^t \frac{\partial}{\partial w}\bar{\Phi}\left[w(x-\varepsilon \Delta(u) )+  u \right]du. 
\end{eqnarray}
With enough accuracy we may substitute the partial derivative~$\partial_{w}$ with~$\partial_{u}$. Integrating by part, taking in account that~$\bar{\Phi}\left[w(x-\varepsilon \Delta(t) )+  t \right]|_{t=0}= \bar{\Phi}(w(x))=\Phi(x,0)~$ we finally obtain

\begin{eqnarray}\label{eigmx_d}
\Phi(x,t)&\approx& \bar{\Phi}\left[w(x-\varepsilon \Delta(t) )+  t \right]. 
\end{eqnarray}
The solution of the linear problem is giving us the solution of the non-linear problem (\ref{liou3}) setting as constant the argument of function~$ \bar{\Phi}(z)$, i.e.

\begin{eqnarray}\label{eigmx_d2}
w(x-\varepsilon \Delta(t) )+  t=\frac{1}{\varepsilon}\int_0^{x-\varepsilon \Delta(t)} \frac{1}{v'(y)-f(t)} dy+t=\mathrm{constant}. 
\end{eqnarray}
The constant can be determined using the initial condition~$x=x_0$ at~$t=0$. It is worthy to stress the following points:
\begin{enumerate}[a)]
\item When~$f(t)$ is a constant Eq.~(\ref{eigmx_d2}) gives the known exact solution.
\item The difficulties to invert Eq.~(\ref{eigmx_d2}) with respect to~$x$ are the same of the exact case, i.e. when~$f(t)=\mathrm{constant}$.
\item The derivative of the potential function~$v'(x)$ is subjected to the less restrictive condition~$|v'(x)|< K$ with~$K$ a positive constant, while~$f(t)$ must satisfy the more strict condition (\ref{condi}).
\end{enumerate}
We now will consider an example that will illustrate the points (a), (b), (c). We will study Eq.~(\ref{eigmx_d2}) using as potential a periodic potential [that does not satisfy (\ref{condi})] in presence of  an arbitrary force~$f(t)$. This problem is known as the diffusion in the egg-carton potential~\cite{caratti,caratti2}. The movement equation is

\begin{eqnarray}\label{liou3_cos}
\frac{dx}{dt}=\varepsilon\left[a \cos(\beta x)+f(t)\right],\,\, x(0)=0.
\end{eqnarray}
According to Eq.~(\ref{eigmx_d2}) the solution of Eq.~(\ref{liou3_cos}) is given by  

\begin{eqnarray}\label{sol_cos}
\frac{2 \tanh ^{-1}\left[\frac{(a-f(t)) \tan \left[\frac{\beta (x-\varepsilon\Delta(t))}{2}\right]}{\sqrt{a^2-f^2(t)}}\right]}{\beta  \sqrt{a^2-f(t)^2}}+\varepsilon t=0
\end{eqnarray}
The inversion of Eq.~(\ref{sol_cos}) is straightforward and~$x(t)$ can be found explicitly. This step is left to the reader. In Fig.~\ref{nonlin} we compare the numerical and the analytical solution in the case of a slowly damped harmonic force~$f(t)$

\begin{eqnarray}\label{force_cos}
f(t)=f_0+\frac{b\cos (\Omega t)}{1+\frac{t^2}{\tau ^2}}.
\end{eqnarray}
The agreement is excellent and the percent error evaluated with formula (\ref{norm}) is smaller than~$1\%$. A detailed study of the solution is out of the scope of this paper.

\section{Conclusion}
 In this paper, it has been shown how to determine two coefficients in the Zassenhaus expansion. The analytical expression of these coefficients allowed a series of applications ranging from~$n$th order linear equations, evaluation of eigenvalues, to first order non-linear equations. We have been able to build an approximate closed formula at first order of the expansion parameter~$\varepsilon$ in the case of linear differential equations and non-linear first order differential equation. We showed the applications regarding the evaluation of eigenvalues and the statistical mechanics. Finally, we gave a formula implying an infinite matrix product of exponential operators for the time evolution of a quantum state. According to the expression of the Hamiltonian, the product of the of exponential operators can be summed and expressed as a closed analytical formula. The presented approach could be extended to evaluate other coefficients in the Zassenhaus expansion and to solve higher order non-linear differential equations. This is left for a future work.
 
\section*{Acknowledgments}
he author acknowledges financial support from UTA Mayor project No. 8765-17



\begin{thebibliography}{bib}

\bibitem {camp}J. E.
Campbell, On a law of combination of operators, Proc. London Math.
Soc., \textbf{29}, 14 (1898). (doi.org/10.1112/plms/s1-29.1.14 )

\bibitem {baker} H. F. Baker, Alternant and continuous groups, Proc. London
Math. Soc. (Second series), \textbf{3}, 24 (1905). (doi.org/10.1112/plms/s2-3.1.24)

\bibitem {hausd}F. Hausdorff, Die
symbolische Exponential Formel in der Gruppen theorie, Leipziger
Ber., \textbf{58}, 19 - 48 (1906).

\bibitem {magnus}W. Magnus, On the exponential solution of differential equations for a linear operator, Comm. Pure Appl. Math., \textbf{7}, 649 - 673 (1954). (doi.org/10.1002/cpa.3160070404 )

\bibitem {que}C. Quesne, Disentangling q-Exponentials: A General Approach, Intern. J. Theo. Phys., \textbf{43}, 545 - 559 (2004). (doi.org/10.1023/B:IJTP.0000028885.42890.f5)

\bibitem {casas}S. Blanes a, F. Casas b, J.A. Oteo c, J. Ros, The Magnus expansion and some of its applications, Physics Reports 470 (2009) 151-238

\bibitem {marini}U. M. B. Marconi  and  F. Corberi, Europhys.  Lett., Time dependent Ginzburg-Landau equation for an N-component model of self-assembled fluids, \textbf{30}, 349 - 354 (1995) (doi.org/10.1209/0295-5075/30/6/006)

\bibitem {kim}Hyeong-Chan Kim, Min-Ho Lee, Jeong-Young Ji, and Jae Kwan Kim, Heisenberg-picture 
approach to the exact quantum motion of a time-dependent forced harmonic oscillator, Phys. Rev. A, 53, 3767 (1996). (doi.org/10.1103/PhysRevA.53.3767)

\bibitem {cerv}J. M. Cerver\'{o} and J. D. Lejarreta, The time-dependent canonical formalism: Generalized harmonic oscillator and the infinite square well with a moving boundary, Europhys. Lett., \textbf{45},6 - 12 (doi.org/10.1209/epl/i1999-00123-2)
(1999).

\bibitem {pedro}I. A. Pedrosa and Alexandre Rosas, Electromagnetic Field Quantization in Time-Dependent Linear Media, Phys. Rev.
Lett. \textbf{103}, 010402-1-010402-4 (2009) (doi.org/10.1103/PhysRevLett.103.010402)

\bibitem {zurek} H. T. Quan and W. H. Zurek, Testing quantum adiabaticity with quench echo, 2010 New J. Phys. 12, 093025 (16pp)(doi:10.1088/1367-2630/12/9/093025)

\bibitem {appl1} Zolt\'an Dar\'azs and Tam\'as Kiss, Time evolution of continuous-time quantum walks on dynamical percolation graphs, J. Phys. A: Math. Theor. 46 (2013) 375305 (doi.org/10.1088/1751-8113/46/37/375305)

\bibitem {bader}Philipp  Bader, Arieh Iserles, Karolina Kropielnicka, Pranav Singh, , Effective Approximation for the Semiclassical Schršdinger Equation, Found. Comput. Math. 14, 689-720, 2014. (doi.org/10.1007/s10208-013-9182-8) 

\bibitem {zass3} Francisco Soto-Eguibar and H. M. Moya-Cessa, Solution of the Schr\"odinger Equation for a Linear
Potential using the Extended Baker-Campbell-Hausdorff Formula, Appl. Math. Inf. Sci. 9, 175-181 (2015).

\bibitem {bader2} Bader P, Iserles A, Kropielnicka K, Singh P.  Efficient methods for linear Schrödinger equation in the semiclassical regime with time-dependent potential. Proc. R. Soc. A 472: 20150733 (18pp), 2016. (doi.org/10.1098/rspa.2015.0733)

\bibitem {zass4} M. A. Soloviev, Dirac's monopole, quaternions, and the Zassenhaus formula, Phys. Rev. D 94, 105021 (10pp) (2016). (doi.org/10.1103/PhysRevD.94.105021)

\bibitem {don}Don S. Lemons, Paul LangevinÕs 1908 paper ÔÔOn the Theory of Brownian MotionÕÕ,  Am. J. Phys. \textbf{65}, 1079 - 1081 (1997). (doi:10.1119/1.18725)

\bibitem {peter} Jing-Dong Bao, Yi-Zhong Zhuo, Fernando A. Oliveira, and Peter H\"anggi, Intermediate dynamics between Newton and Langevin, Phys. Rev. E  \textbf{74}, 061111 (2006). (doi.org/10.1103/PhysRevE.74.061111)

\bibitem {van}N. G. Van Kampen, Stochastic differential equations, Phys. Rep. 24, 171 - 228  1976.

\bibitem {west_app}B J West, On dissipative nonlinear Hamiltonian systems,
Applied Stochastic Processes, 283-301, Academic Press (1980) (doi.org/10.1016/B978-0-12-044380-2.50017-6)

\bibitem {west_coll2} M. Bianucci, R. Mannella, P. Grigolini, B.J. West, The Linear Response Approach to the Fokker-Planck Equation I: Theory, Int. J. Mod. Phys. B 8, 1191-1210 (1994). (doi.org/10.1142/S0217979294000567)

\bibitem {west_coll3}M. Bianucci, R. Mannella, P. Grigolini, B.J. West, The Linear Response Approach to the Fokker-Planck Equation II: a Nonlinear Stochastic Booster, Int. J. Mod. Phys. B 8, 1211-1224 (1994). (doi.org/10.1142/S0217979294000579)

\bibitem {west_coll4}M. Bianucci, R. Mannella, P. Grigolini, B.J. West, The Linear Response Approach to the Fokker-Planck Equation III: a Deterministic and Chaotic Booster, Int. J. Mod. Phys. B 8, 1225-1246 (1994). (doi.org/10.1142/S0217979294000580)

\bibitem {pet3}L. Wang, N. Li, and P. H\"anggi
\emph{Thermal transport in low dimensions: From statistical physics to nanoscale heat transfer}, S. Lepri, ed.
Lecture Notes in Physics, vol. 921, pp. 239-274, Springer-Verlag, Berlin, Heidelberg, New York (2016).

\bibitem {bruce}B.J. West, Fractional Calculus View of Complexity: Tomorrow's Science (CRC Press, 2016).

\bibitem {gos}MalgorzataTuralska, Bruce J.West,  A search for a spectral technique to solve nonlinear fractional differential equations
Chaos, Solitons \& Fractals, 102, 387 (2017). (doi.org/10.1016/j.chaos.2017.04.022)

\bibitem {yo} M. Bologna, Exact analytical approach to differential equations with variable coefficients, Eur. Phys. J. Plus (2016) 131: 386 (11pp) (doi.org/10.1140/epjp/i2016-16386-9)

\bibitem {zass1m}Michael Weyrauch and Daniel Scholz, Computing the Baker-Campbell-Hausdorff series and the Zassenhaus product, Computer Physics Communications 180 (2009) 1558 - 1565 (doi.org/10.1016/j.cpc.2009.04.007)

\bibitem {zass2m}Fernando Casas, Ander Murua, Mladen Nadinic, Efficient computation of the Zassenhaus formula, Computer Physics Communications 183 (2012) 2386 - 2391 (doi.org/10.1016/j.cpc.2012.06.006)

\bibitem {zass1} 
J\"urgen Geiser, Gamze Tano$\check{\mathrm{g}}$lu, Nurcan G\"uc\"uyenen, Higher order operator splitting methods via Zassenhaus product formula: Theory and applications, Computers \& Mathematics with Applications, 62, 1994 - 2015 (doi.org/10.1016/j.camwa.2011.06.043)

\bibitem {nelson}E. Nelson, Feynman integral and the Schršdinger equation, J. Math. Phys. \textbf{5}, 332 - 343 (1964) (doi.org/10.1063/1.1704124)

\bibitem {landau3}L. D. Landau and E. M. Lifshitz Quantum Mechanics Non-relativistic 
Theory, 3rd Ed., Pergamon Press (1991).

\bibitem {landau4}L. D. Landau and E. M. Lifshitz Quantum Electrodynamics , 2ed, Pergamon Press (1982).

\bibitem {kelley}W.G. Kelley, A.C. Peterson. \emph{The Theory of Differential Equations},  Springer (2010).

\bibitem {4} J.-P. Bouchaud, A. Georges, Anomalous diffusion in disordered media: statistical mechanisms, models and physical applications, Phys. Rep. 195, 127-293 (1990)  

\bibitem {5} R. Metzler, J. Klafter, The random walk's guide to anomalous diffusion: a fractional dynamics approach, Phys. Rep. 339, 1-77 (2000) (doi.org/10.1016/S0370-1573(00)00070-3)

\bibitem {6} G.M. Zaslavsky, Chaos, fractional kinetics, and anomalous transport, Phys. Rep. 371, 461-580 (2002) (doi.org/10.1016/S0370-1573(02)00331-9)

\bibitem {7} Mauro Bologna and Gerardo Aquino, Weakly driven anomalous diffusion in non-ergodic regime: an analytical solution, Eur. Phys. J. B (2014) 87: 15 (7pp) (doi.org/10.1140/epjb/e2013-40701-3)

\bibitem {pet}J. Spiechowicz, J. Luczka, and P. H\"anggi
Transient anomalous diffusion in periodic systems: ergodicity, symmetry breaking and velocity relaxation 
Scientific Reports 6, 30948 (11pp) (2016). (doi.org/10.1038/srep30948)

\bibitem {sancho2}J. M. Sancho  M. San Miguel Some Results in the Description of Systems under the Influence of Dichotomous Noise 
Progr. Theoret. Phys. 69, 1983, 1085-1090 (doi.org/10.1143/PTP.69.1085)

\bibitem {sancho} J.M. Sancho, Stochastic processes driven by dichotomous Markov noise: Some exact dynamical results, J. Math. Phys. \textbf{25}, 354 (1984) (doi.org/10.1063/1.526160)

\bibitem {caratti}G. Caratti, R. Ferrando, R. Spadacini and G. E. Tommei, Physica A: Statistical Mechanics and its Applications,  
An analytical approximation to the diffusion coefficient in overdamped multidimensional systems, 246, 1997, 115-131. (doi.org/10.1016/S0378-4371(97)00345-2)

\bibitem {caratti2}G. Caratti, R. Ferrando, R. Spadacini and G. E. Tommei, Underdamped diffusion in the egg-carton potential, Phys. Rev. E 55, 1997, 4810. (doi.org/10.1103/PhysRevE.55.4810)



\end{thebibliography}
\end{document}